\documentclass[twocolumn]{aastex7}

\usepackage{xspace}
\usepackage{hyperref}
\usepackage{ulem}
\usepackage{placeins}
\usepackage[utf8]{inputenc}

\newcommand{\asec}{$^{\prime\prime}$\xspace}

\newcommand{\NII}{\mbox{[N\,\textsc{ii}]}}

\shorttitle{Star formation in M87}
\shortauthors{Tamhane et al.}

\begin{document}

\title{Constraining star formation in M87 using deep HST UV data}

\correspondingauthor{Prathamesh Tamhane, Ming Sun}
\email{pdt0003@uah.edu, ms0071@uah.edu}

\author[orcid=0000-0001-8176-7665]{Prathamesh Tamhane}
\affiliation{Department of Physics and Astronomy, University of Alabama in Huntsville, 301 Sparkman Drive, Huntsville, AL 35899, USA}
\email{pdt0003@uah.edu}

\author[orcid=0000-0002-6009-0197]{William Waldron}
\affiliation{Department of Physics and Astronomy, University of Alabama in Huntsville, 301 Sparkman Drive, Huntsville, AL 35899, USA}
\email{wvw0330@uah.edu}

\author[orcid=0000-0001-5880-0703]{Ming Sun}
\affiliation{Department of Physics and Astronomy, University of Alabama in Huntsville, 301 Sparkman Drive, Huntsville, AL 35899, USA}
\email{ms0071@uah.edu}

\author[orcid=0000-0001-7110-6775]{Silvia Martocchia}
\affiliation{Aix Marseille University, CNRS, CNES, LAM, Marseille, France}
\email{silvia.martocchia@lam.fr}

\author[orcid=0000-0001-7711-3677]{Claudia Maraston}
\affiliation{Institute of Cosmology and Gravitation, University of Portsmouth, 1-8 Burnaby Road, Portsmouth, PO1 3FX, UK}
\email{Claudia.Maraston@port.ac.uk}

\author[orcid=0000-0002-9795-6433]{Alessandro Boselli}
\affiliation{Aix Marseille University, CNRS, CNES, LAM, Marseille, France}
\affiliation{Osservatorio Astronomico di Cagliari, Via della Scienza 5, 09047 Selargius (CA), Italy}
\email{alessandro.boselli@lam.fr}

\author[orcid=0000-0002-9478-1682]{William Forman}
\affiliation{Center for Astrophysics $|$ Harvard \& Smithsonian, 60 Garden Street, Cambridge, MA 02138, USA}
\email{wforman@cfa.harvard.edu}

\author[orcid=0000-0003-2754-9258]{Massimo Gaspari}
\affiliation{Department of Physics, Informatics and Mathematics, University of Modena and Reggio Emilia, Modena, Italy}
\email{massimo.gaspari@unimore.it}

\author[orcid=0000-0002-7705-4483]{Juhi Tiwari}
\affiliation{Department of Physics and Astronomy, University of Alabama in Huntsville, 301 Sparkman Drive, Huntsville, AL 35899, USA}
\email{jt0189@uah.edu}

\author[orcid=0000-0002-2808-0853]{Megan Donahue}
\affiliation{Michigan State University, Physics and Astronomy Department, East Lansing, MI 48824-2320, USA}
\email{donahu42@msu.edu}

\author[orcid=0000-0002-3514-0383]{G. Mark Voit}
\affiliation{Michigan State University, Physics and Astronomy Department, East Lansing, MI 48824-2320, USA}
\email{voit@pa.msu.edu}

\author{Tim Edge}
\affiliation{Department of Physics and Astronomy, University of Alabama in Huntsville, 301 Sparkman Drive, Huntsville, AL 35899, USA}
\email{edge.timothy00@gmail.com}

\author[orcid=0000-0002-5445-5401]{Grant Tremblay}
\affiliation{Center for Astrophysics $|$ Harvard \& Smithsonian, 60 Garden Street, Cambridge, MA 02138, USA}
\email{grant.tremblay@cfa.harvard.edu}

\author[orcid=0000-0002-6325-5671]{Daniel Thomas}
\affiliation{Institute of Cosmology and Gravitation, University of Portsmouth, 1-8 Burnaby Road, Portsmouth, PO1 3FX, UK}
\email{daniel.thomas@port.ac.uk}



\begin{abstract}

We analyzed the deepest Hubble Space Telescope (HST) F275W ultraviolet (UV) imaging of M87 to obtain the most robust constraints on its star formation rate (SFR) and star formation history (SFH). After removing the galaxy continuum and globular clusters, we detected an excess of UV point sources near the center. By comparing their colors to young stellar source (YSS) colors generated by stochastically simulated star formation (SF) for various SFRs and SFHs, we ruled out their origin as a UV-upturn population and identified them as YSS. We found an extremely low SFR of $\sim 2\times10^{-5}$ M$_\odot$ yr$^{-1}$ in M87, with evidence of a weak starburst $\sim$125 Myr ago that formed $\sim 1000$ M$_\odot$ of stars. Unlike other cool-core clusters where SF is stronger and directly linked to cooling gas, we found no spatial correlation between YSS and H$\alpha$ filaments. Comparing SF activity with M87's AGN outburst history suggests that recent AGN feedback events ($\lesssim$12 Myr ago) neither triggered nor were associated with any detectable SF, however, earlier outbursts may have triggered weak starbursts. We detected UV filaments co-spatial with H$\alpha$ filaments with similar lengths and widths, though they are obscured by dust near the center. These filaments are likely powered by metal-line emission from collisional ionization, suggesting ongoing low-level precipitation of the intracluster medium. Our results indicate that AGN feedback has quenched SF significantly in M87 for at least 200 Myr, even though some precipitation persists. Additionally, we identified a hotspot created by the counterjet, with the spectral index also constrained.

\end{abstract}



\section{Introduction} 
\label{sec:intro}

Multiphase filaments of cold and warm gas are prevalent in the central galaxies of cool core clusters, appearing in over one-third of all galaxy clusters and groups \citep[e.g.,][]{mcnamara12,gaspari20,dv22}. These filaments—traced by CO \citep{salome03, russell19,olivares19,tamhane22}, atomic gas \citep{edge10a}, dust \citep{rawle12}, warm H$_2$ \citep{donahue11}, and H$\alpha$ emission \citep{crawford99, hamer16}—are clear evidence of ongoing cooling from the intracluster medium (ICM), despite the presence of radio-mode AGN feedback. Yet, even in systems with abundant cold gas, star formation is surprisingly weak, typically only $\sim 1$–2\% of the classical X-ray cooling rate \citep[e.g.,][]{mcdonald18}. This low star formation efficiency challenges our understanding of the cooling-feedback cycle and BCG growth.

Star formation is the final sink in the cooling process and depends on the balance of gravity, cooling, magnetic fields, turbulence, and bulk flows in the cool core. While strong radio AGN are known to regulate cooling and quench star formation, they may also briefly trigger it \citep[e.g.,][]{silk13,salome15,tamhane23}. Intriguingly, nearly all systems with high SFRs ($>10$ M$_\odot$ yr$^{-1}$) are radio-loud, but many radio-loud cool cores exhibit only weak star formation. Given short AGN duty cycles ($\sim$10–-100 Myr; \citealt{shulevski15}), it remains unclear whether strong AGN quickly quench starbursts or whether all such systems experience brief episodes of enhanced star formation.

To investigate this, a detailed study on nearby X-ray cool cores with strong radio AGN but weak SFR is required. We build a complete sample of nearby, strong X-ray cool cores from \citet{Sun09a} and \citet{cavagnolo09}, with these criteria: a) $z < 0.02$; b) ($L_X > 10^{42} \, \text{erg s}^{-1}$) for the cool core; c) system $kT >$ 2 keV to only study clusters. There are only four such clusters, Virgo (M87), Centaurus, AWM7, and Perseus (NGC 1275). Among them, only NGC 1275 shows a high SFR ($\sim 20$ M$_\odot$ yr$^{-1}$; \citealt{canning10}), consistent with the $\sim 1/3$ fraction of cluster BCGs with SFRs above 5 M$_\odot$ yr$^{-1}$ \citep{mcdonald18}. M87, in contrast, hosts a strong radio AGN and a bright filamentary nebula but has an SFR of just $\sim 0.2$–0.3 M$_\odot$ yr$^{-1}$ \citep{hoffer12}. It is also closest to us making it an ideal target for studying suppressed star formation in the presence of multiphase gas.

M87 is the BCG of the nearby Virgo cluster (16.5 Mpc; \citealt{mei07}) and one of the best-studied radio galaxies. Its cool core, with a central cooling time $<$1 Gyr within 10 kpc \citep{cavagnolo09}, shows X-ray cavities, shocks, and cold fronts from recurrent AGN outbursts \citep[e.g.,][]{forman05, werner10, forman17}. Optical and UV observations reveal a network of emission-line filaments and $10^5$ K gas \citep{sparks12, boselli19, anderson18}, likely formed through uplift of low-entropy  gas and thermal instabilities in the ICM \citep{werner13}. While ALMA detects no central CO \citep{boizelle25}, molecular gas and dust are present in outer filaments \citep{alighieri13,werner13,simionescu18}, confirming the multiphase nature of the system.

Though star formation in M87 is weak, existing very deep HST UV imaging of M87 combined with its proximity allow an unprecedented detailed study of its star formation activity and history. Section~\ref{sec:data} describes the HST data and analysis; Section~\ref{sec:pt_src_removal} covers point source detection; Section~\ref{sec:sfr} presents star formation modeling; Section~\ref{filaments} discusses UV-bright filaments; Section~\ref{discussion} provides interpretation, and Section~\ref{sec:conclusion} concludes the paper.
At M87's distance (16.5 Mpc), 1\asec = 80 pc.

\begin{table*}
    \begin{center}
    \caption{HST data used in this study}
    \label{tab:obs}
    \vspace{-0.3cm}
    \begin{tabular}{cccccccc}
    \hline
     Filter & Instrument & Proposal ID & Obs Year & N$^{\rm a}$ & Tot Exp (ksec) & Mean $\lambda$ (\AA) & Area (arcmin$^2$) \\
    \hline
        F275W  & WFC3/UVIS & 12989, 14618, 16894 & 2013, 2017--2018, 2022 & 185 & 110.11 & 2719 & 11.98 \\
        F606W  & WFC3/UVIS & 14618, 16894 & 2017--2018, 2022 & 112 & 40.04 & 5999 & 11.98 \\
        F660N  & ACS/WFC   & 12271 & 2011 & 18 & 13.88 & 6599 & 2.12 \\
        F814W  & ACS/WFC   & 10543, 13731 & 2005--2006 & 268 & 87.21 & 8129 & 12.37 \\
    \hline
    \end{tabular}
    \end{center}
    \vspace{-0.2cm}
    Note: $^{\rm a}$: the number of exposures.
\end{table*}

\section{HST observations and Data Analysis}
\label{sec:data}

The HST data used in this work were collected with Advanced Camera for Surveys (ACS) and Wide Field Camera 3 (WFC3) instruments focusing on the inner~10\,kpc
of the galaxy. Observations from multiple filters (F275W, F606W, F660N-Pol, and F814W)
provide complementary insights, with the F275W filter highlighting recent star formation, the F660N-Pol filter revealing the optical emission-line nebula, and the F606W and F814W filters showing the evolved stellar population. 
Together, these filters enable the
estimation of stellar ages within the target galaxy and its surroundings. The data used in
this paper came from the following proposals --- 10543 (PI: Baltz), 12271 (PI: Sparks), 12989 (PI: Renzini), 13731 (PI: Meyer), 14618 (PI: Shara), and 16894 (PI: Neilsen), as summarized in Table~\ref{tab:obs}. Combined, this field is one of the deepest F275W fields with HST.

\begin{figure*}
    \includegraphics[width=\textwidth]{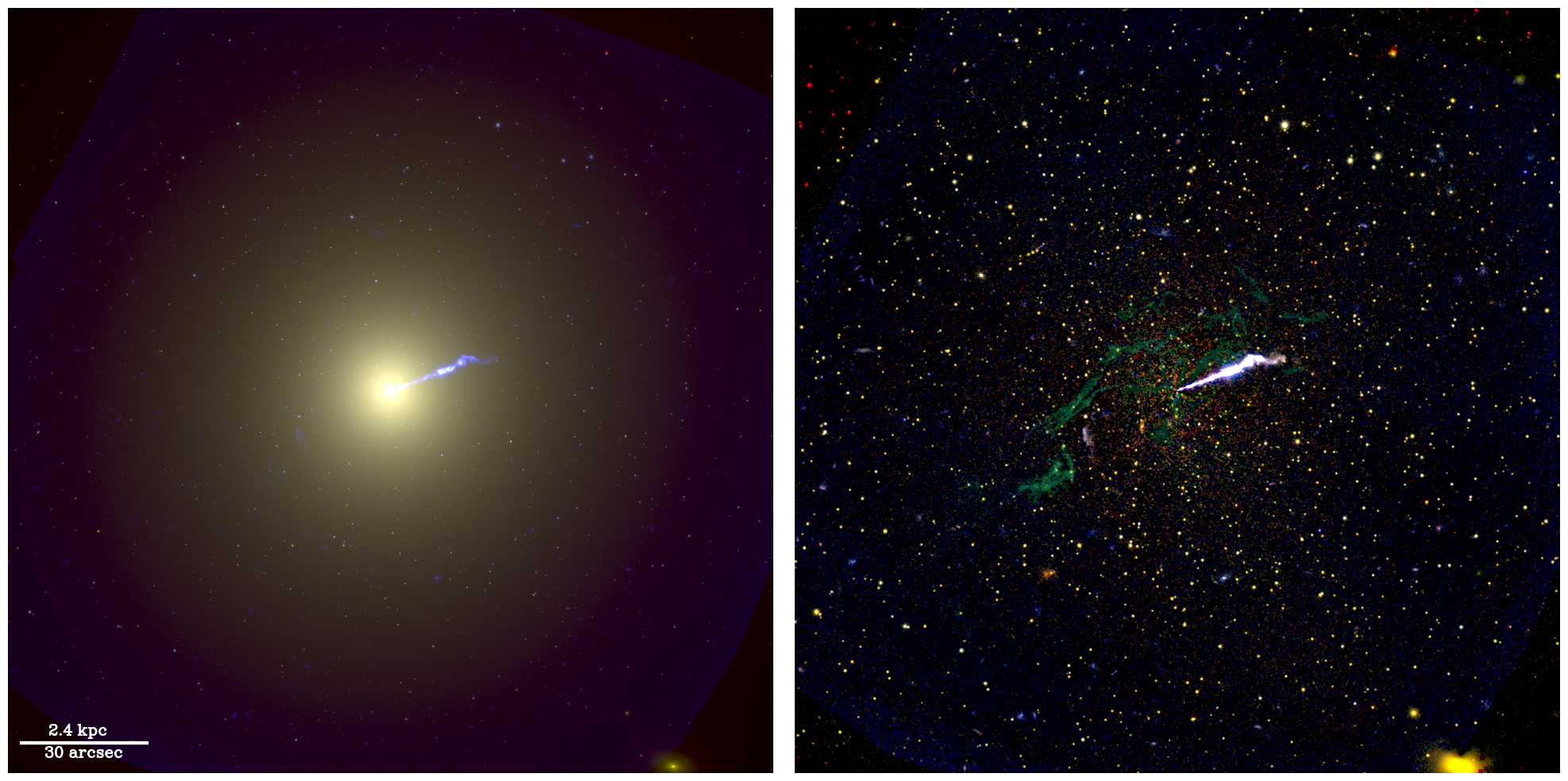}
    \caption{{\em Left panel:} Composite RGB image of M87 (blue: WFC3 F275W, green: F606W, red: ACS F814W). {\em Right panel:} The same image after subtracting the diffuse galactic emission and further smoothed using a Gaussian kernel with a standard deviation of 4 pixels. H$\alpha$ emission from HST ACS F660N image is shown in Green. This image highlights compact sources and H$\alpha$ filaments otherwise obscured by the underlying diffuse light. A few blue spiral galaxies are visible in the outskirts of the image, whereas most of the yellow point sources are M87 globular clusters. North is up and East is to the left in both panels.
    }
    \label{fig:rgb}
\end{figure*}

To prepare the data for photometric analysis, the images were aligned to the GAIA~Data Release 3
catalog \citep{gaia2016, gaia2023} to within 0\farcs{}01 using the
\textsc{TweakReg} tool from the \textsc{DrizzlePac} package. 
For data with a small field of view, 
the HST pipeline drizzled images from each dither position were used in alignment, and the final alignment was back-propagated to the input images using \textsc{TweakBack} within the \textsc{DrizzlePac}.
Images in the same band taken on different dates were photometrically normalized using
\textsc{PhotEq} in the \textsc{DrizzlePac}.
Subsequently, the images in each band were drizzled together using \textsc{AstroDrizzle},
which reduced noise, removed cosmic rays in overlapping regions, and produced a
consistent pixel scale of 0\farcs{}03 for all bands.
The output weight images were set to inverse variance maps to provide accurate noise
estimates for photometry using Source Extractor (SExtractor). Additionally, root mean square (RMS) images were generated by taking the inverse square root of the inverse variance maps, allowing for proper uncertainty estimates in the photometric analysis. This pipeline ensured that the data were processed uniformly and optimized for robust photometric measurements in the current study. The final images have a 5$\sigma$ limiting magnitudes of~28.21\,mag,~28.01\,mag, and~28.78\,mag in F275W, F606W and F814W bands, respectively, in 0$''$.5 diameter circular aperture.
Figure~\ref{fig:rgb} shows the RGB image made by combining F275W (Blue),
F606W (Green) and F814W (Red) images after subtracting the diffuse galactic emission
modeled using \textsc{Photutils}\footnote{\url{https://photutils.readthedocs.io/en/stable/}} isophote subtraction (see Sec.~\ref{sec:pt_src_removal}) and smoothed with a gaussian kernel of 4 pix.
All our data products are available at MAST as a High Level Science Product via \dataset[10.17909/ 1yjm-c412]{\doi{10.17909/1yjm-c412}}.

\section{Point source detection}
\label{sec:pt_src_removal}
We performed point source identification on the F275W image using SExtractor, following the subtraction of diffuse galaxy emission. The diffuse emission was removed by iterative isophotal fitting implemented with \textsc{Photutils} following the example notebooks\footnote{\url{https://github.com/astropy/photutils-datasets/tree/main/notebooks/isophote}}. For regions not covered by the elliptical fits, we applied 2D background subtraction using photutils with a 50$\times$50 pixel box size to remove large-scale residuals.

\begin{table*}
    \centering
    \caption{Summary of positions and magnitudes of UV sources detected in this study. Magnitudes are corrected for the Galactic extinction.}
    \begin{tabular}{c c c c c}
        \hline
        RA & DEC & F275 & F606 & F814 \\
        (h:m:s) & (d:m:s) & (ABmag) & (ABmag) & (ABmag) \\
        \hline
        12:30:49.46 & 12:21:44.88 & 24.551$\pm$0.075 & 20.877$\pm$0.002 & 20.437$\pm$0.001 \\
        12:30:51.29 & 12:21:45.98 & 23.649$\pm$0.039 & 20.776$\pm$0.002 & 20.452$\pm$0.001 \\
        12:30:51.88 & 12:21:47.12 & 24.944$\pm$0.157 & 24.219$\pm$0.058 & 24.716$\pm$0.011 \\
        12:30:47.64 & 12:21:47.24 & 25.807$\pm$0.111 & 23.009$\pm$0.007 & 22.596$\pm$0.005 \\
        12:30:47.38 & 12:21:47.85 & 24.731$\pm$0.039 & 17.316$\pm$0.000 & 16.647$\pm$0.005 \\
        \hline
    \end{tabular}
    \tablecomments{Only the first five sources are shown here; the full table is available in the online version.}
    \label{tab:src_summary}
\end{table*}

SExtractor operates by first estimating the background across the image, typically using a mesh grid and median filtering to account for large-scale background variations. It then detects sources by identifying contiguous groups of pixels above a specified threshold relative to the background. We adopted a detection threshold of 3$\sigma$. Additionally, the image was smoothed using a 5$\times$5 pixel Gaussian kernel with a standard deviation of 2.5 pixels to enhance the detectability of faint sources. Photometric fluxes were measured in a 0$''$.5 diameter circular aperture centered on each detected source. The aperture size is sufficiently large to detect individual stars and young star clusters, which typically range in size from a few to 20 pc. We performed multi-band photometry using SExtractor's dual-image mode. Source detection was performed on the F275W image, while photometry was measured on the F606W and F814W images. All images were aligned to ensure consistent astrometry, with errors less than 0$''$.1.

We masked the central AGN and its jet to avoid contamination from bright structures. After source detection, we identified 1384 sources. Table~\ref{tab:src_summary} summarizes source properties. We also compared our sources to the classical novae detected in M87 by \cite{shara23} and \cite{lessing24}. We cross-matched our source catalog with their nova catalog using a matching radius of 0$''$.07, identifying 28 common sources across the entire image, which were subsequently removed from our UV source catalog. This radius was chosen to account for astrometric uncertainties while minimizing false matches. Finally, we selected 1219 sources with a signal-to-noise ratio greater than 5 for further analysis. The measured source fluxes were corrected for Galactic extinction using $A_V = 0.0612$ taken from \citet{schlafly11} dust maps and \citet{fitzpatrick99} extinction curve with $R_V = 3.1$ and converted to AB magnitudes.

\begin{figure*}
    \centering
    \includegraphics[width=\textwidth]{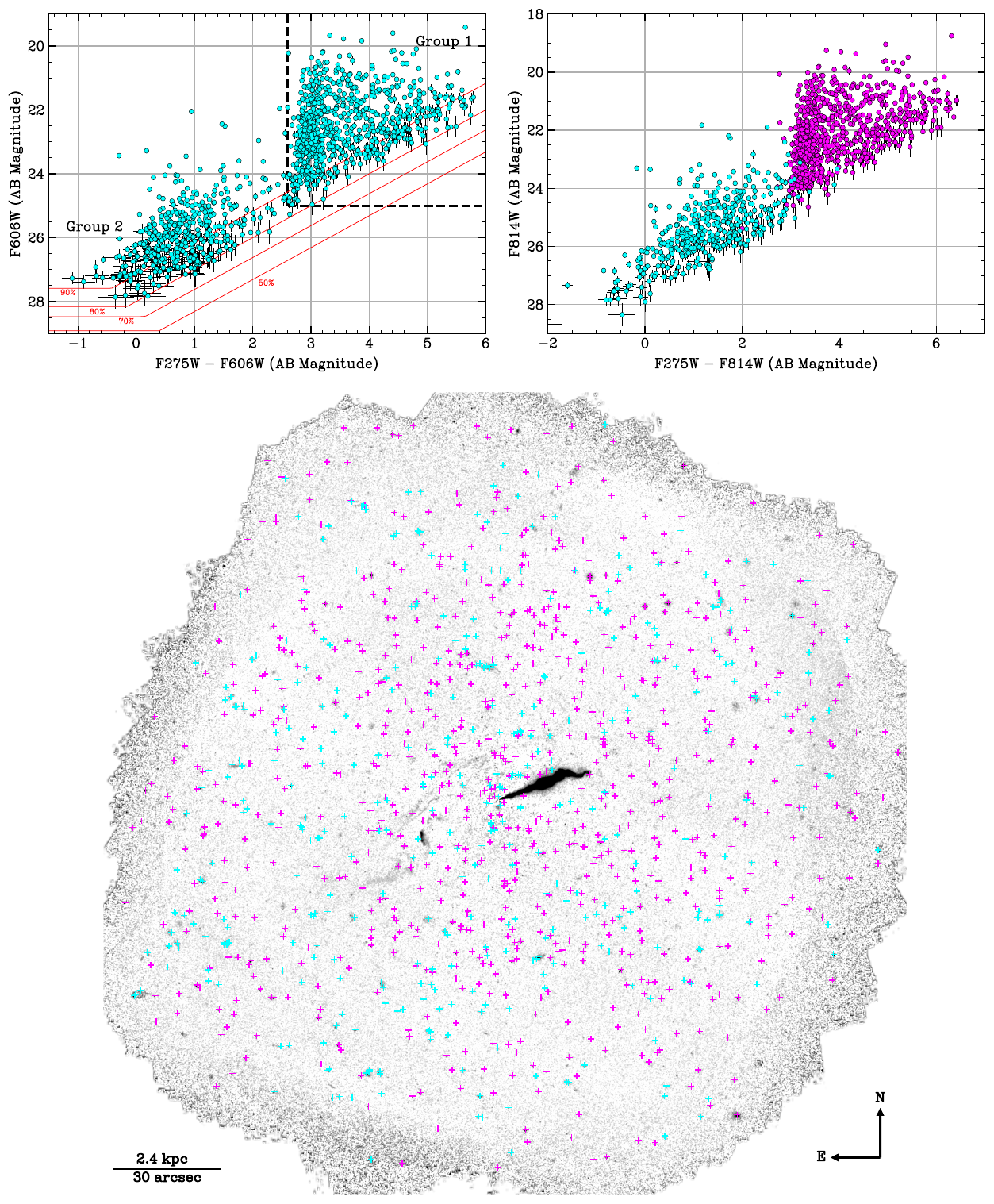}
    \caption{{\em Top Left:} Color-magnitude relation of all point sources detected in HST WFC3 F275W and F606W images of M87, showing a clear separation into two distinct groups. The black dashed lines denote the boundary separating these groups. The red lines show the completeness percentage.
    {\em Top Right:} Color-magnitude diagram for F275W and F814W band magnitudes. Group 1 points defined in the left panel are shown in Magenta color. The two clusters are separated in this CDM as well, however, the separation between the two groups is smaller. {\em Bottom:} The F275W residual image after subtracting galaxy continuum emission shows the detected UV sources and filaments. The Magenta and Cyan crosses are ``Group 1'' and ``Group 2'' sources, respectively.}
    \label{fig:cm}
\end{figure*}

Figure~\ref{fig:cm} shows the color-magnitude diagram (CMD) for all point sources detected in M87 in F275W image. In the left panel, sources are distinctly grouped based on F275W$–$F606W color and F606W magnitude, forming two main populations. We categorize sources with F275W$–$F606W $>$ 2.6 mag and F606W $<$ 25 mag as Group 1 and the remaining sources as Group 2. There are 803 sources in Group 1 and 416 sources in Group 2. Sources in Group 1 with redder colors, likely correspond to globular clusters (GCs) in M87, as previously characterized by \citet{bellini15}. However, \citet{bellini15} focused specifically on GCs and used only 26\% of the F275W data analyzed here. Our deeper UV data enabled the detection of a bluer population in Group 2, consistent with young stellar sources. Many sources in this group have F606W magnitudes in the range of 24--28.
The right panel shows the F275W$-$F814W color and F814W magnitude. The two populations of sources are visible in this CMD as well; however, they appear closer together. The bottom panel of Figure~\ref{fig:cm} shows the spatial distribution of sources in M87. The magnitudes of our Group 1 sources are consistent with those of the globular clusters in \citet{bellini15} to within 0.006 mag in all bands.

\pagebreak

Since distinguishing galaxy-associated and background sources is challenging, we used the Hubble Deep UV (HDUV) Legacy Survey\footnote{\url{https://archive.stsci.edu/prepds/hduv/}} to estimate background source density. The HDUV (GOODS-N/S) fields offer comparable background conditions due to their depth, high Galactic latitude, and use of the same HST filter. We applied the same source detection and photometry method as for the M87 field for consistency, which is SExtractor with a 3$\sigma$ detection threshold, 50$\times$50 pixel background box, and a 5$\times$5 pixel Gaussian filter.

To evaluate source detection reliability, we conducted completeness tests on both M87 and HDUV images by injecting artificial point sources (FWHM 4--5 px, F275W mag 20--31.5) at random positions and re-running SExtractor. The completeness level was quantified as the ratio of the number of detected artificial sources to the total number of injected sources. M87 image completeness is 94\% for sources with mag $<$26.5, 91\% at 27.5, and 80\% at 28.5. In the HDUV images, completeness is 100\% to mag 23.5, 90\% at 27, and 35\% at 28.5.

\begin{figure*}
    \centering
    \includegraphics[width=0.45\textwidth]{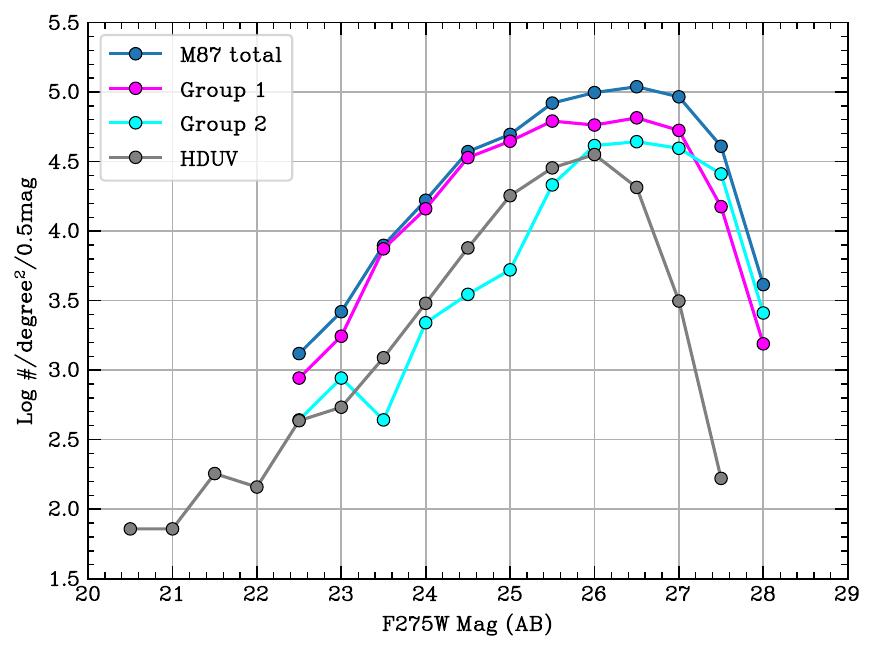}
    \includegraphics[width=0.45\textwidth]{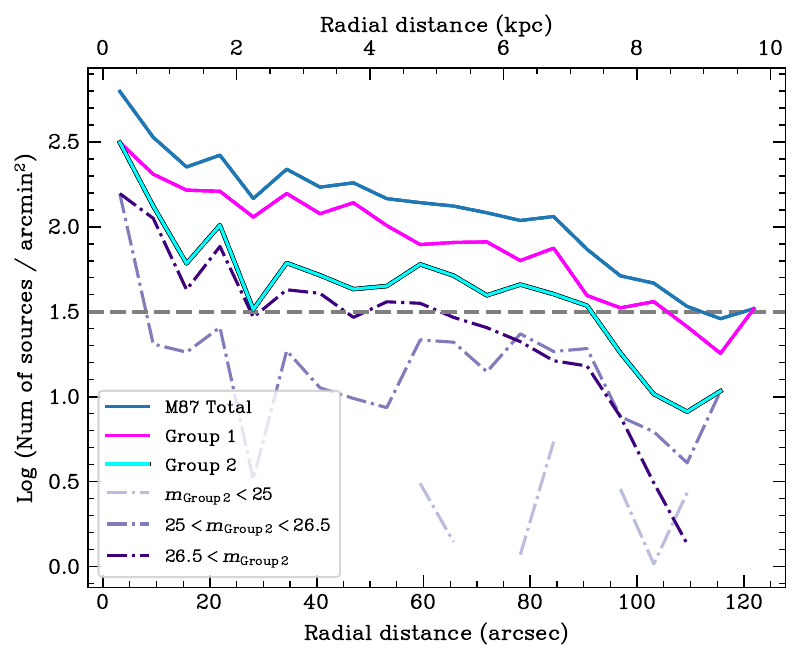}
    \caption{\emph{Left Panel:} The number density of point sources in HST F275W images of M87 and HDUV fields per 0.5 mag per square degree plotted for comparison. \emph{Right panel:} Radial distribution of point sources in M87 F275W image from the center of the galaxy. The purple dash-dotted lines show the radial profile of Group 2 sources in different magnitude bins.
    }
    \label{fig:src_num_den}
\end{figure*}

In the left panel of Figure~\ref{fig:src_num_den}, we present the source surface density for both the M87 image and the HDUV fields. The source number densities in the figure are adjusted to account for completeness. The total source number density in M87 is 10 or more times higher than in the HDUV field in every magnitude bin where both data are present. In the HDUV fields, the average total source surface number density is $\sim$ 31 sources per arcmin$^2$. Thus the expected number of background sources in the M87 image is $\sim$ 276 which is $\sim$ 22\% of the total detected point sources in the F275W image in M87. The right panel of Figure~\ref{fig:src_num_den} shows the radial distribution of source surface number density in logarithmic units as a function of distance from the galaxy's center without completeness correction. It increases towards the galaxy's center by approximately $\sim$1 dex for all sources and for sources in Group 1. The Group 2 source number density increases from $\sim$50 arcmin$^{-2}$ at $\sim40''$ to $\sim$320 arcmin$^{-2}$ in the innermost region. This trend indicates an excess of blue sources in the inner region of M87, where most of the increase is in the inner 30$''$ region. There are $\sim$192 sources in this region of which only $\sim$24 sources are expected to be background sources, which is only $\sim$12.5\% of the sources. This trend clearly indicates that there is an excess of UV-bright sources in the inner region of the galaxy.
There are 67 sources in Group 2 in the inner 30$''$ region with approximately $\sim 12$\% expected to be background sources. This still leaves approximately 58 Group 2 sources that we identify as young stellar sources.

We also estimated the contribution of older stars to the diffuse galactic UV light, known as the UV-upturn \citep[e.g.,][]{Martocchia25}. The expected F275W$-$F606W color of UV-upturn sources in M87, from the \cite{Martocchia25} model, is significantly redder ($\sim$ 5.15) compared to the much bluer Group 2 sources ($-1$ to 2.6), confirming that UV-upturn stars do not contribute to the Group 2 population. In Appendix~\ref{sec:rad_prof}, we discuss the galaxy's light and color profiles in detail and compare them with the expected UV emission from older stars. Additionally, we compared the colors of blue horizontal branch stars at high metallicity with ages of 10 and 15 Gyr from the \citet{maraston05} model and found that they have F275W$-$F606W colors of 4.17 and 4.30, respectively. Thus, these stars are bluer than UV-upturn-emitting stars but redder than the YSCs detected in M87.

\begin{figure*}
    \includegraphics[width=\textwidth]{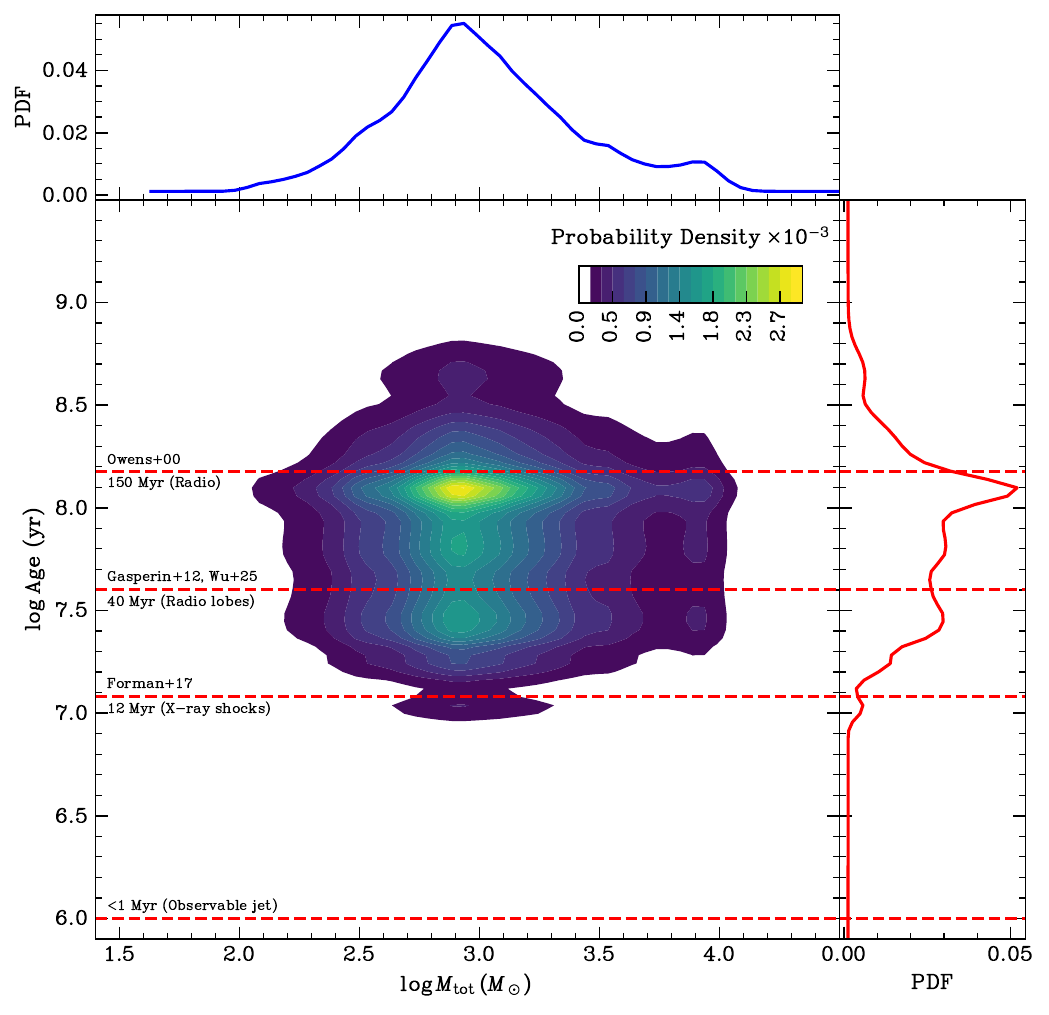}
    \caption{The figure shows the 2D posterior probability density distribution of cluster age and mass, with density indicated by the colorbar. To the top and right of the main panel, we show the marginalized PDFs of cluster mass and age, respectively. The dashed red lines mark the times of past AGN outbursts inferred from X-ray cavity ages and radio lobes.}
    \label{fig:slug_result}
\end{figure*}

\section{Star formation Rate in M87}
\label{sec:sfr}
One of the ways to measure the star formation rate and history is to compare observations to simulations of simple stellar populations in a galaxy with known SFR and SFH. Since star formation in M87 is expected to be very low, we use the code Stochastically Lighting Up Galaxies (SLUG)\footnote{\href{https://bitbucket.org/krumholz/slug2}{https://bitbucket.org/krumholz/slug2}} \citep{dasilva12,krumholz15}. The SLUG code is a stellar population synthesis tool that models star clusters and galaxies by stochastically sampling individual stars and clusters, rather than assuming a continuous initial mass function (IMF). This approach allows the code to capture the effects of stochasticity in young, low-mass star clusters, providing more accurate predictions in the low star formation regime, which traditional synthesis codes based on averaged populations often miss.

For an initial analysis, we ran SLUG simulations in galaxy mode, modeling both continuous and single-burst SFHs with Chabrier IMF \citep{chabrier03}, Padova isochrones including pulsating AGB stars, and metallicities of 0.05 and 0.02. For each SFH, we simulated 100 trials using a range of parameters to estimate the expected distribution of star cluster properties under different star formation scenarios. For continuous star formation, we applied a timescale of 200 Myr and SFRs between $10^{-4}$ and $10^{-6}$ M$_\odot$ yr$^{-1}$. For single-burst SFH, we assumed an exponential decay over 10 Myr, with burst peaks ranging from 6 Myr to 200 Myr and total stellar masses from 10$^2$ to $10^4$ M$_\odot$. The simulation generated 35–50 star clusters per trial, consistent with the expected number of Group 2 sources in the inner 30$''$ region, with mean cluster masses between 30--40 M$_\odot$, and HST band magnitudes within the observed range. To determine the best model producing simulated star cluster distribution closely resembling the observed distribution of `Group 2' points, we computed the Bhattacharyya distance between simulated clusters of each parameter set and the observed sources. A Bhattacharyya distance of 0 represents identical source distributions with perfect overlap and larger values indicate diverging distributions with increasingly smaller overlap \citep{bhattacharyya46}. We found that
a single burst SFH model, a burst 125 Myr ago with 10$^3$ M$_\odot$ stars formed for solar ($Z=0.02$) metallicity tracks gave the best match between observed and simulated sources with a Bhattacharyya distance of 0.17. It implies a time-averaged SFR rate of $\sim 10^{-5}$ M$_\odot$ yr$^{-1}$.

For a more detailed analysis, we used the forward modeling approach presented in \citet{krumholz19} to infer the underlying mass and age distribution of the observed star cluster population. We used their extensive library of simulated star clusters with a Chabrier IMF and solar metallicity, spanning a wide range of cluster ages and masses. Rather than simulating new cluster populations for each assumed SFH, we reweighted each cluster by adjusting its relative probability based on its age to reflect the assumed SFH. We assumed a uniform prior on cluster age, allowing the data to constrain the SFH shape without bias. We then compared the reweighted simulated cluster sample to the observed HST magnitudes using a likelihood function, following the method described in \citet{krumholz19}. The resulting posterior probability density distribution is shown in Figure~\ref{fig:slug_result}. We sampled the posterior probability distribution functions (PDFs) of stellar mass and age 10$^5$ times to infer the median mass, age, and SFR with 1$\sigma$ credible intervals. The total stellar mass formed is $995^{+3446}_{-573}$ M$_\odot$, with a cluster age of $7.1^{+10.0}_{-4.8} \times 10^7$ yr, corresponding to a time-averaged SFR of $1.8^{+1.7}_{-0.4} \times 10^{-5}$ M$_\odot$ yr$^{-1}$. The PDF of cluster age shows a clear peak at $\sim 125$ Myr and drops steeply at younger ages, approaching zero at $\sim 8$ Myr, indicating little to no ongoing star formation. The distribution also shows minor secondary peaks, including one around $\sim 28$ Myr, which coincides with the estimated age of M87's outer radio lobe \citep{gasperin12,wu25}, suggesting a possible link between AGN activity and star formation. The overall shape of the PDF supports a scenario of burst-like or episodic star formation rather than continuous activity.

\section{UV bright filaments}
\label{filaments}

M87 hosts a complex network of filaments extending several kiloparsecs from the center, observed across X-ray, UV, optical and IR wavelengths \citep[e.g.,][]{sparks93, sparks09, werner10, werner13, forman17, boselli19}. These filaments are also detected in the HST WFC3 F275W and F606W images and are co-spatial with the H$\alpha$ filaments. The left panel of Figure~\ref{fig:filaments} shows the filaments detected in the F275W image after subtracting a smooth galaxy model. The right panel shows the similarly processed HST ACS F660N image containing the H$\alpha$ emission, where a 2D galaxy model has been subtracted (as described in Section~\ref{sec:pt_src_removal}). In both images, arrows highlight the outer filaments, visible in F275W and H$\alpha$. We also detected a bright feature in the F275W image, located $\sim 2$ kpc south-east of the nucleus, indicated by cyan arrow in Fig.~\ref{fig:filaments}. This feature is not detected in the F665N image and coincides with the bright radio emission associated with M87's counterjet. We interpret this as a hotspot created by the interaction of the radio jet with the surrounding medium.

All filaments detected in F275W are close but appear disconnected from each other. The filaments have lengths between 0.7 to 1.2 kpc and FWHM widths $\sim 22$ pc to 50 pc. The N filament appears thicker with a $\sim$ 70 pc width. However, it may be composed of several narrow thread-like filaments as seen in the F665N image in the right panel, where nebular filaments appear twisted. The eastern and southeastern filaments are more thread-like. The two sub-filaments in the E filament are $\sim$ 236 pc (2.9$''$) apart, whereas the narrow subfilaments in the SE filament are $\sim$ 400 pc apart. All of these filaments are co-spatial with the H$\alpha$+\NII{} filaments shown in the right panel of Fig~\ref{fig:filaments}. The lengths and widths of UV filaments are consistent with H$\alpha$+\NII{} filaments. The actual lengths can be longer depending on the relative orientation of the filaments to the line of sight.

The visibility of the outer H$\alpha$ filaments in the HST F275W images, in contrast to the non-detection of the inner H$\alpha$ filaments, is explained by differences in dust extinction. Multiple H$\alpha$ filaments stretch from the nucleus to the north and south, but the inner filaments are not detected in the F275W image. These inner filaments overlap with regions of significant dust absorption, as revealed by the dust absorption contours in the figure obtained from the unsharp-masked F606W image.

We estimated the extinction in these regions in the regions covered by the absorption contours as $A_{606} = -2.5 \, {\rm Log((F606_{mod}-abs(F606_{res}))/F606_{mod})}$, where $\rm F606_{mod}$ and $\rm F606_{res}$ are fluxes extracted from the model image of the galaxy's continuum and residual image, respectively, within dust absorption regions. $A_V$ and expected $A_{275}$ were calculated using the \citet{cardelli89} extinction law and $R_V = 3.1$. We found an average visual extinction $A_V = 0.015 \pm 0.008$, consistent with previous studies. The expected extinction in the F275W band is $\sim 0.034$, while the observed extinction in F275W, estimated in those same regions following a similar approach as that for F606W image, is $A_{275} = 0.031 \pm 0.028$, which is close to the expected value. This level of extinction is sufficient to obscure the inner H$\alpha$+\NII{} filaments in the F275W image. The flux ratio between the F275W regions that are cospatial with the bright inner H$\alpha$+\NII{} filaments (in the original data, before subtracting the galaxy continuum) and the regions at the same radial distance that are not cospatial with the filaments is $\sim 2.9 \pm 0.6$ \%.
The outer filaments, however, show little to no extinction along the line of sight, which explains their visibility.

\section{Discussion}
\label{discussion}

\subsection{Star formation history and AGN feedback}

\begin{figure*}
    \centering
    \includegraphics[width=\textwidth]{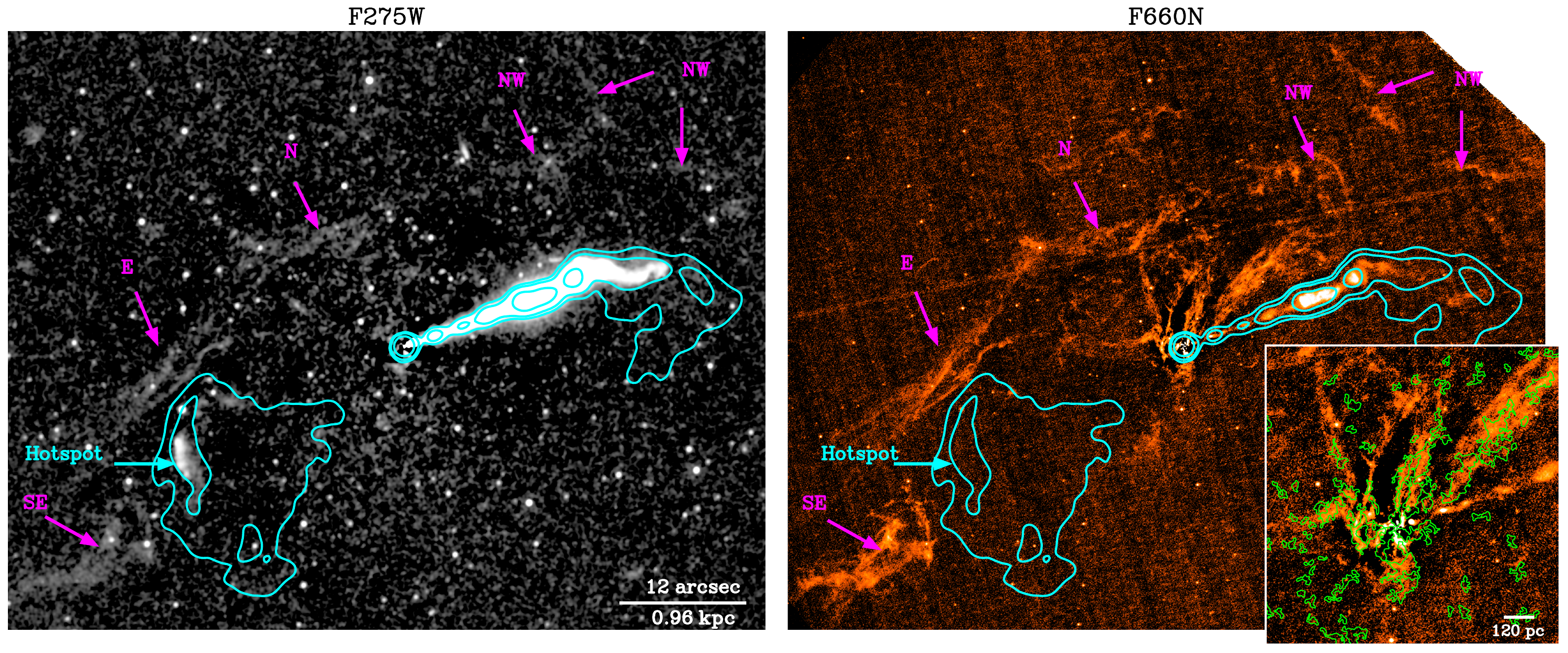}
    \caption{Residuals in the WFC3 F275W image (left) after subtracting a galaxy continuum model and the ACS F660N image containing the H$\alpha$+\NII{} emission (right). The cyan contours in both images represent Very Large Array (VLA) 1.4 GHz radio emission at levels of 0.12, 0.19, 0.59 and 2.84 Jy/beam. The inset image in the right panel provides a zoomed-in view of the nucleus in F660N, highlighting H$\alpha$+\NII{} filaments overlapping with dust absorption contours (green). The magenta arrows indicate UV and optical filaments, while the cyan arrow shows the hotspot created by the counterjet, which is detected in the F275W image but not in F660N. North is up in both images.
    }
    \label{fig:filaments}
\end{figure*}

Early studies predicted a continuous SFR of up to 1 M$_\odot$ yr$^{-1}$ throughout the lifetime of 1 Gyr of the inner cooling flow in M87 \citep{fabian84}. \citet{hoffer12} reported M87's SFR between 0.2--0.3 M$_\odot$ yr$^{-1}$ based on integrated IR luminosity of the galaxy and an upper limit of 0.1 M$_\odot$ yr$^{-1}$ from GALEX UV flux upper limit. They used SFR-IR scaling relations from \citet{calzetti10} to convert IR fluxes to SFR, which have a scatter of 0.2 dex and are less reliable in the low SFR regime. This estimate is also likely an overestimate because the integrated IR flux is dominated by synchrotron emission from the AGN \citep{boselli10,baes10} which was not taken into account and subtracted from the total flux. In this work, we measured the most robust estimate of the current SFR and SFH in the last 200 Myr in M87 using the deepest HST UV data available. The star formation rate in M87 in the last $\sim 200$ Myr is very low at $\sim10^{-5}$ M$_\odot$ yr$^{-1}$ with a very weak starburst around 125 Myr ago leading to the formation of stars with a total stellar mass of $\sim 1000$ M$_\odot$. No supernovae have been reported in M87 over the last $\sim$ 100 years, which also supports the conclusion that its current star formation rate is low.

M87 has experienced at least three AGN outbursts in the past 200 Myr. The jet corresponds to the current outburst, estimated to be less than 1 Myr old. \citet{forman17} identified a previous outburst 11--12 Myr ago based on X-ray shocks. The absence of stars with comparable ages suggests that these recent outbursts did not trigger new star formation, consistent with the general picture of AGN feedback in massive early type galaxies, where mechanical energy from jets and shocks suppresses cooling and prevents cold gas accumulation \citep[e.g.,][]{mcnamara07, fabian12}. However, the youngest stars observed in M87's inner regions have ages comparable to the AGN outburst that occurred approximately 100--150 Myr ago, inferred from the extended radio halo \citep{owen00}, as shown in Fig.~\ref{fig:slug_result}. This suggests that this earlier outburst may have triggered a mild starburst. However, updated age estimates of the outer halo place its age between 30--50 Myr \citep{gasperin12,wu25}. The small bump in the stellar age PDF at $\sim$30 Myr may correspond to this event. If so, the main peak may correspond to an even earlier AGN outburst, the signatures of which have since faded or are undetectable, or to another mechanism entirely.

The presence of H$\alpha$ filaments and a small amount of molecular gas ($\sim 10^5$ M$_\odot$, \citealt{simionescu18}) suggests that localized cooling persists, but not at levels sufficient to sustain substantial star formation. Despite M87's short central ICM cooling time ($10^7$–$10^8$ yrs; \citealt{russell15}) and a classical cooling rate of $\sim$20 M$_\odot$ yr$^{-1}$ \citep{russell13}, XMM RGS measurements show much lower cooling rates, $<0.06$ M$_\odot$ yr$^{-1}$ below 0.5 keV \citep{werner10} and 0.8 M$_\odot$ yr$^{-1}$ for an absorbed cooling flow \citep{fabian23}, although the RGS coverage does not fully include the cool core.

Star formation in M87 remains highly inefficient, likely due to a combination of mechanical heating, weak residual cooling, and uplift of low-entropy gas. Past AGN outbursts may have uplifted a significant amount of low-entropy gas to larger radii, reducing the supply of cool gas in the central regions \citep{werner10, mcnamara16}. Since cold gas is the primary fuel for star formation, its depletion may have led to suppression of star formation in M87. AGN-driven turbulence can also disrupt cooling, preventing the gas from settling into a star forming phase. Sometimes it can lead to sporadic outbursts of low-level star formation and temporally correlated AGN outbursts \citep{voit18, gaspari17}. Nonetheless, some level of residual cooling likely persists and may intermittently fuel the central SMBH.

While AGN feedback has largely suppressed star formation in M87, the presence of H$\alpha$ and UV filaments, despite the disruptive effects of the AGN, raises the question of what mechanisms are responsible for their excitation. Understanding how these filaments remain ionized, particularly in the absence of significant star formation, provides crucial insights into the interplay between cooling, feedback, and the surrounding ICM.

\subsection{Excitation mechanisms of the filaments in the F275W Band}
Although we lack spectroscopic data in the F275W band to directly determine the mechanism powering the filaments at these wavelengths, insights can be drawn from previous studies in the FUV. The Mg \textsc{ii} $\lambda2797$, $\lambda2803$ \AA~doublet would be covered by the F275W filter. The doublet traces neutral gas and can be strong in star forming regions and AGN \citep[e.g.,][]{guseva13,popovic19}. To assess whether ionizing radiation from young stars could explain the observed F275W flux, we modeled a starburst spectrum for a 1000 M$_\odot$ stellar population. Our calculations show that reproducing the UV filament flux would require stars younger than 20 Myr with a total mass of approximately 500--1000 M$_\odot$. However, the lack of young UV sources in the galaxy strongly suggests that the filaments are not photoionized by stars. The observed star formation around the filaments is approximately 100 times lower than what is required to power them. Spectroscopic observations with HST COS of the E filament revealed that it is powered by line emission from collisional excitation, including lines such as C \textsc{iv}, He \textsc{ii}, C \textsc{ii}, Ly$\alpha$, and N \textsc{v} \citep{sparks12, anderson18}. Such emission lines, some of which correspond to $10^5$ -- $10^6$ K gas, are expected if the ICM gas is cooling from $\sim10^7$ K to $10^4$ K or even lower temperatures \citep[e.g.,][]{fabian22}. Similar processes likely contribute to the observed brightness of the outer filaments in this band. Simulation of a hidden cooling flow \citep[e.g.,][]{fabian22} with CLOUDY suggests that the filament may be dominated by Fe and Mg emission lines around 2550--2800 \AA~(private communication).

We also estimated the contribution of shock ionization by two-photon emission, a process often observed in supernova shocks that can produce near- and far-UV continuum emission. Using the framework and parameters outlined in Appendix B of \citet{bracco20}, we estimate the two-photon emission flux in the F275W band to be $\sim$ 3--6 $\times10^{-20}$ erg s$^{-1}$ cm$^{-2}$ \AA$^{-1}$ - three orders of magnitude lower than the observed filament flux ($\sim$ 0.8--2 $\times10^{-17}$ erg s$^{-1}$ cm$^{-2}$ \AA$^{-1}$). This suggests that two-photon emission is negligible in the filaments and that the observed flux is instead dominated by metal-line cooling from collisionally excited gas. Collisional excitation by the surrounding ICM, as well as mixing processes, may also contribute \citep[e.g.,][]{canning16}.

\subsection{Connection between UV sources and optical/UV filaments}
Figure~\ref{fig:cm} shows the spatial distribution of all detected sources in the galaxy. Visual inspection reveals no apparent correlation between the sources and the optical H$\alpha$ filaments. This contrasts with NGC 1275, where young star clusters are closely associated with optical filaments, strongly suggesting that star formation in NGC 1275 is recent and fueled by cooling gas in cold filaments \citep{canning14}. While the filaments in M87 may have formed through the thermally unstable cooling of hot X-ray gas in the ICM, some studies suggest that they could also be remnants of a merger or interaction with a star-forming galaxy \citep{boselli19}. Nevertheless, if stars in M87 formed from gas in the filaments, a spatial correlation between the filaments and young stars would be expected. The absence of such a correlation suggests that the young stars in M87 are not associated with the optical filaments. On the other hand, stars decouple from the gas as soon as they form and move ballistically. Therefore, even a velocity difference as small as 10 km s$^{-1}$ between the gas and stars could result in a separation of 1.5 kpc (18$''$) or more over the lifetime of stars making such spatial correlation difficult to observe if present.

The reason for the absence of star formation in these filaments is not well understood. \citet{anderson18} propose that the low atomic gas mass in M87's filaments, compared to those in NGC 1275 and similar systems, may be a contributing factor. If the filaments form through cooling of the ICM, their increasing density may enhance the magnetic field strength \citep[e.g.,][]{fabian16}. The narrow width of the filaments suggests strong magnetic fields. \citet{werner13} estimated magnetic field strengths of 30--70 $\mu$G in these filaments, whereas equipartition arguments suggest higher values of $\gtrsim$100 $\mu$G (Tamhane et al., in prep). AGN jets may also induce turbulence in the surrounding medium \citep{wittor20}, and combined magnetic and turbulent pressure could suppress gravitational collapse, suppressing star formation \citep{federrath12}. Additionally, if the filaments are young and still accumulating mass, they may not yet have reached the threshold for star formation. Based on radial velocities from Far-UV lines and their distance from the nucleus, these filaments may be just a few 10$^7$ years old and are likely dynamic systems evolving in mass and temperature \citep{anderson18}. Thus, their young age, low gas mass, and strong magnetic and turbulent support may all contribute to the suppression of star formation.

\subsection{Counterjet hotspot}
\label{hotspot}

We detected a hotspot produced by the counterjet in the F275W image, as shown in the left panel of Fig.~\ref{fig:filaments}. It is 3--4 times brighter than the filaments and is also visible in the F606W and F814W images. Additionally, we inspected the HST F150LP image from the HST Legacy Archive and found that the hotspot is detected in the far-UV (FUV) as well. We present its spectral energy distribution (SED) using available data.

We measured the hotspot flux within an elliptical aperture with semi-major and semi-minor axes of 2$''$.32 and 0$''$.81, respectively, and a position angle of 12 degrees, ensuring consistent coverage across all HST. We used the galaxy continuum subtracted HST images in the F275W, F606W and F814W bands. For F150LP, we subtracted the local background from an annular region around the hotpot excluding bright sources. The extracted HST fluxes were corrected for Galactic extinction in a similar way as for UV sources. We used the same region to extract flux from the VLA and ALMA images of the hotspot. VLA images were obtained from the NRAO VLA Archive Survey Images page\footnote{\url{http://www.vla.nrao.edu/astro/nvas/}} and ALMA image was obtained from the pipeline processed product from the ALMA data archive with project ID 2013.1.00862.S (PI: Simionescu). No background subtraction was performed for VLA and ALMA data; therefore, the hotspot flux density may include some contribution from the radio lobe. We also used $\sim$ 1 Ms of Chandra data from 46 ACIS observations to constrain the X-ray emission for the hotspot, following standard Chandra ACIS data processing steps and measuring fluxes in the 0.5--1.5 keV and 5--9 keV bands assuming an absorbed powerlaw model with a photon index of 1.7. 
The hotspot is not detected. 
All extracted flux densities or limits are listed in Table~\ref{tab:hotspot}.

\begin{figure}
    \centering
    \includegraphics[width=0.45\textwidth]{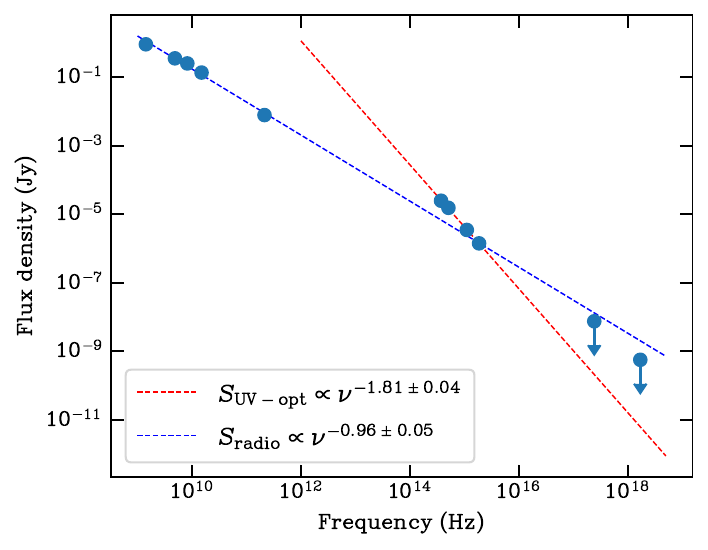}
    \caption{The SED of M87's counterjet hotspot. The dashed lines are the best fits for a powerlaw spectrum.
    }
    \label{fig:hotspot}
\end{figure}

Since the hotspot emission is powered by synchrotron radiation \citep[][hereafter S92]{sparks92}, we calculated its spectral index $\alpha$ assuming a power-law spectrum $S \propto \nu^{-\alpha}$, where $S$ is the flux density and $\nu$ is the frequency. The optical-to-UV spectral index derived from only HST data is $1.81 \pm 0.04$, which is lower than the previously reported value of $2.23 \pm 0.5$ from S92 using ground-based optical data for the average hotspot emission, as shown by the red line in Fig.~\ref{fig:hotspot}. The radio spectral index from 1.4 to 214 GHz is $0.96 \pm 0.05$, much shallower than the optical spectral index, indicating a break in the synchrotron spectrum at frequencies between 214 GHz and 372 THz. We did not perform the detailed modeling of the hotspot SED as it is beyond the scope of this paper. As discussed by S92, the light travel time from the nucleus to the hotspot ($\sim 10^4$ years) exceeds the synchrotron lifetime of electrons in the hotspot (a few hundred years), implying that continuous energy injection is required to sustain the emission.

\begin{table}[]
    \centering
    \begin{tabular}{ccc}
    \hline
        Band & Freq (Hz) & Flux density (Jy) \\
        \hline\hline
        VLA L & 1.42$\times 10^9$ & 0.92$\pm$0.03 \\
        VLA C & 4.86$\times 10^9$ & (3.58$\pm$0.004)$\times 10^{-1}$ \\
        VLA X & 8.21$\times 10^9$ & (2.54$\pm$0.01)$\times 10^{-1}$ \\
        VLA U & 14.96$\times 10^9$ & (1.38$\pm$0.02)$\times 10^{-1}$ \\
        ALMA & 214.1$\times 10^9$ & (7.90$\pm$0.61)$\times 10^{-3}$ \\
        HST F814W & 3.73$\times 10^{14}$ & (24.94$\pm$0.01)$\times 10^{-6}$ \\
        HST F606W & 5.09$\times 10^{14}$ & (15.46$\pm$0.02)$\times 10^{-6}$ \\
        HST F275W & 1.11$\times 10^{15}$ & (3.50$\pm$0.02)$\times 10^{-6}$ \\
        HST F150LP & 1.86$\times 10^{15}$ & (1.42$\pm$0.03)$\times 10^{-6}$ \\
        Chandra 1 keV & 2.42$\times 10^{17}$ & $<7.52 \times 10^{-9}$ \\
        Chandra 7 keV & 1.69$\times 10^{18}$ & $<5.64 \times 10^{-10}$ \\
        \hline
    \end{tabular}
    \caption{Summary of flux measurement of the hotspot region from HST, ALMA, VLA and Chandra data. The errors are statistical only.
   }
    \label{tab:hotspot}
\end{table}

\section{Conclusion}
\label{sec:conclusion}

We analyzed the deepest HST F275W UV imaging of M87 to constrain its star formation rate and history. The unprecedented sensitivity ($>$28 mag) and spatial resolution ($\sim$ 2.5 pc) enabled the detection of individual star clusters. After subtracting the galaxy continuum and removing contributions from globular clusters, novae, and background sources, we identified an excess of UV sources within the central $\sim$30$''$. Using SLUG models, we constrained their ages and masses. We also detected bright, narrow UV filaments co-spatial with H$\alpha$ and multiwavelength filaments.

Our main findings are summarized as follows:
\begin{itemize}
    \item The time-averaged SFR in M87 is extremely low, $1.8^{+1.7}_{-0.4} \times 10^{-5}$ M$_\odot$ yr$^{-1}$, over four orders of magnitude below the X-ray cooling rate. The age probability distribution peaks around $\sim$125 Myr, consistent with a mild starburst that formed $995^{+3446}_{-573}$ M$_\odot$ of stars.
    \item Comparison of AGN feedback activity with the star formation history suggests that the most recent AGN outbursts in M87 have not led to significant star formation. However, a small bump in the stellar age distribution around $\sim$30 Myr may indicate a minor episode of star formation potentially triggered by a previous AGN outburst, while the main burst at $\sim$125 Myr may have been triggered by an even earlier AGN outburst.
    \item UV filaments are dust-obscured near the center ($A_V \sim 0.015$) and not powered by young stars or two-photon emission. Instead, they likely trace cooling ICM gas, emitting via metal-line cooling from collisional ionization or mixing.
    \item We presented an SED of the hotspot created by the counterjet of M87. Its optical-UV spectral index $1.81 \pm 0.04$, whereas the radio spectral index is $0.96 \pm 0.05$, indicating a break in the synchrotron spectrum.
\end{itemize}

The extremely low star formation rate in M87 suggests that the radio AGN is efficient in quenching star formation in M87. The temporal overlap between the starburst and an AGN outburst suggests that it can occasionally trigger brief episodes of star formation. However, the modest amount of stars formed suggests that these starbursts are transient and do not lead to sustained star formation. Therefore, such brief starburst phases might be a common feature of the feedback cycle in radio-loud X-ray cool cores. Repeated occurrences of such bursts over cosmic time may contribute to the buildup of low-mass stars in the galaxy's core. Our results therefore raise the possibility that this mode of star formation favors low-mass star formation \citep[e.g.,][]{jura77, fabian24}, potentially contributing to the bottom-heavy IMFs observed in massive ellipticals, including M87 \citep[e.g.,][]{vandokkum10, oldham18}.

Future deep multiwavelength observations will deepen our understanding of the AGN feedback in M87. Deep James Webb Space Telescope observations in the near-to-far infrared may reveal dust obscured star formation in the center if present and help characterize properties of stellar populations better, while high-resolution UV spectroscopy will help characterize the ionization state of the gas in the filaments. Moreover, X-ray Imaging and Spectroscopy Mission (XRISM) observations of M87 will constrain the ICM turbulence, providing insights on its role in AGN feedback cycle and star formation in cool-core clusters. Additionally, it will help linking the dynamics of the hot ICM to the cold filaments if they are indeed cooling from the hot ICM.

\begin{acknowledgments}
We thank Mark Krumholz for the help on the SLUG code.
We thank Andy Fabian and Norbet Werner for comments.
Based on observations with the NASA/ESA Hubble Space Telescope obtained from the Data Archive at the Space Telescope Science Institute, which is operated by the Association of Universities for Research in Astronomy, Incorporated, under NASA contract NAS5-26555. Support for Program number (17037) was provided through a grant from the STScI under NASA contract NAS5-26555.
W.F. acknowledges support from the Smithsonian Institution, the Chandra High Resolution Camera Project through NASA contract NAS8-0306, NASA Grant 80NSSC19K0116 and Chandra Grant GO1-22132X.
M.G. acknowledges support from the ERC Consolidator Grant \textit{BlackHoleWeather} (101086804).
Some/all of the data presented in this article were obtained from the Mikulski Archive for Space Telescopes (MAST) at the Space Telescope Science Institute. The specific observations analyzed can be accessed via \dataset[doi: 10.17909/h7yr-0231]{https://doi.org/10.17909/h7yr-0231}. This work has made use of data from the European Space Agency (ESA) mission
{\it Gaia} (\url{https://www.cosmos.esa.int/gaia}), processed by the {\it Gaia}
Data Processing and Analysis Consortium (DPAC,
\url{https://www.cosmos.esa.int/web/gaia/dpac/consortium}). Funding for the DPAC
has been provided by national institutions, in particular the institutions
participating in the {\it Gaia} Multilateral Agreement.
\end{acknowledgments}

%

\vspace{5mm}
\facilities{HST(WFC3), HST(ACS)}


\software{astropy \citep{2013A&A...558A..33A,2018AJ....156..123A,astropy22},  
          Source Extractor \citep{1996A&AS..117..393B},
          SLUG \citep{krumholz15},
          photutils \citep{photutils}
          }



\appendix

\section{Radial galaxy light and color profiles}
\label{sec:rad_prof}
\begin{figure*}
    \centering
    \includegraphics[width=0.47\textwidth]{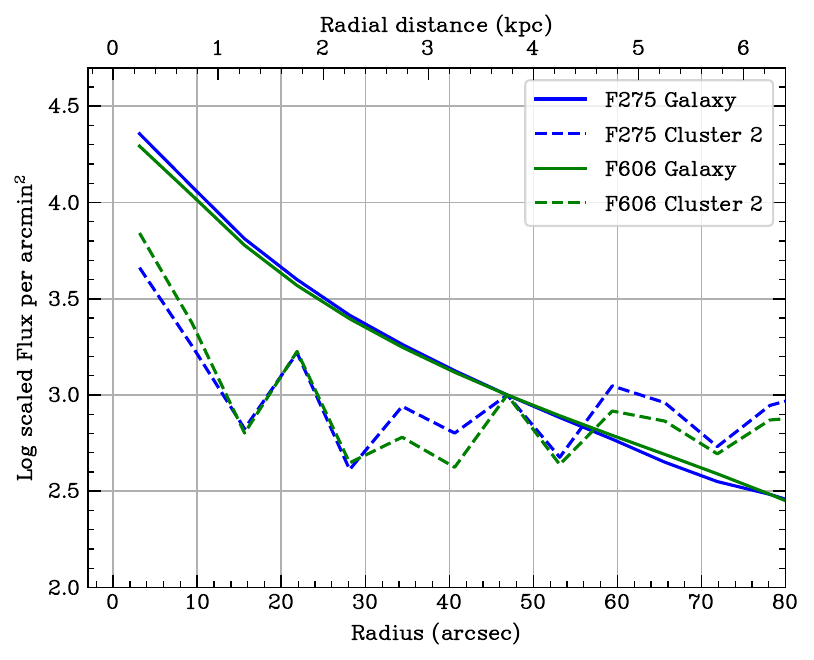}
    \includegraphics[width=0.47\textwidth]{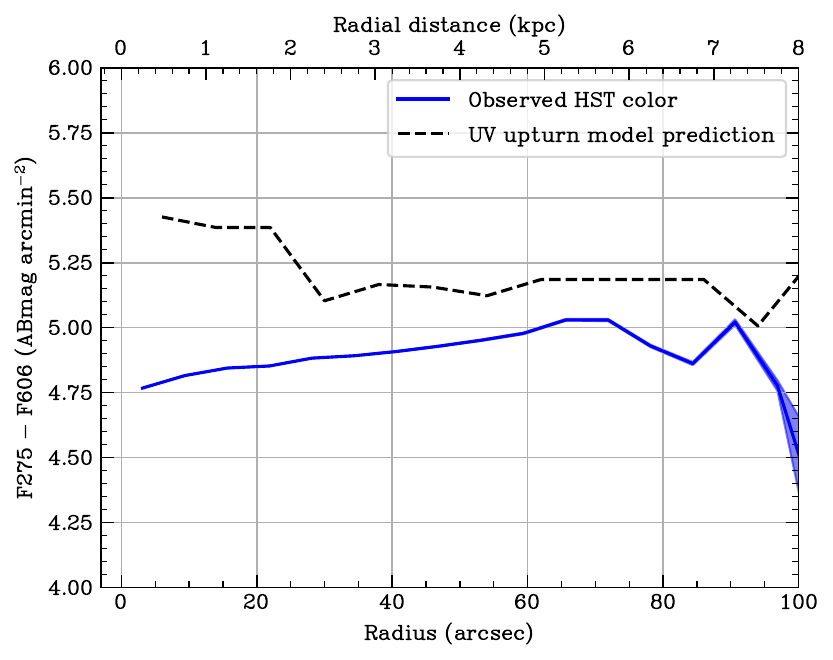}
    \caption{{\em Left panel:} Logarithmic light profiles of galaxy diffuse light and detected point sources as a function of radius in F275W and F606W bands. All fluxes are scaled to the F275W flux at $\sim 53''$. Errors are not included. {\em Right panel:} F275W$-$F606W color profile of M87 in blue derived from our HST data. The shaded blue region shows 1$\sigma$ statistical errors. The dashed black line shows the color profile of the UV-upturn stellar population expected in M87 from \cite{Martocchia25}.
    }
    \label{fig:rad_prof}
\end{figure*}

We estimated radial F275W and F606W light profiles for M87 in the inner $\sim$7 kpc region. To do this, we used both a galaxy model generated with \textsc{Photutils} isophotes and direct measurements from the original image, after masking detected sources, the AGN, and the jet. The resulting profiles are shown in the left panel of Fig.~\ref{fig:rad_prof}. We found negligible differences between the profiles generated from the model image and those derived directly from the original image; therefore, we present the profiles from the original image. For reference, we also scaled the flux from all detected sources in both bands and overlaid them on the galaxy light profiles. Additionally, we derived the F275W$-$F606W color profile of the galaxy and compared it with UV-upturn models. The black dashed line in the right panel of Fig.~\ref{fig:rad_prof} represents the expected color profile of old stars emitting in the UV convolved with M87's SFH from \citet{Martocchia25}.

Our analysis shows that M87 has a mild radial UV flux gradient, similar to its optical flux gradient. The observed F275W$-$F606W color profile gradually reddens from 4.75 at the center to 5.03 at $\sim$5.6 kpc (70$''$). This trend is slightly bluer than predictions from the UV upturn model. The UV upturn model of \cite{Martocchia25} predicts a shallower F275W$-$F606W gradient, with colors changing from 5.43 at the center to 5.15 at 2.4 kpc (30$''$), remaining relatively flat out to 8 kpc (100$''$). These models can produce bluer colors, such as 4.75 for a 12 Gyr old single stellar population with half solar metallicity. However, the central regions of M87 have super solar metallicity \citep{Martocchia25}, which leads to a redder UV upturn color in the core, consistent with the expected metallicity dependence of the UV upturn phenomenon.

\bibliography{M87}{}
\bibliographystyle{aasjournalv7}



\end{document}